\documentclass[twocolumn]{aastex631}
\usepackage{textcomp}
\usepackage{amssymb}
\usepackage{rotating}
\usepackage{threeparttable}

\newcommand{\fermi}{\textit{Fermi}}
\newcommand{\gr}{$\gamma$-ray}
\newcommand{\hawc}{J0631+107}
\newcommand{\lasso}{J0631+1040}

\shorttitle{Two candidate pulsar TeV halos}
\shortauthors{Zheng \& Wang}

\begin{document}

\title{Two Candidate Pulsar TeV Halos Identified from Property-Similarity Studies}


\author{Dong Zheng}
\affiliation{Department of Astronomy, School of Physics and Astronomy, Yunnan University, Kunming 650091, China; zhengdong@mail.ynu.edu.cn; wangzx20@ynu.edu.cn}

\author[0000-0003-1984-3852]{Zhongxiang Wang}
\affiliation{Department of Astronomy, School of Physics and Astronomy,
Yunnan University, Kunming 650091, China; zhengdong@mail.ynu.edu.cn; wangzx20@ynu.edu.cn}
\affiliation{Key Laboratory for Research in Galaxies and Cosmology,
Shanghai Astronomical Observatory, Chinese Academy of Sciences, 
80 Nandan Road, Shanghai 200030, China}


\begin{abstract}
	TeV halos have been suggested as a common phenomenon associated with
	middle-aged pulsars. Based on our recent work on the middle-aged
	pulsar J0631+1036, which is the only known source positionally
	coincident with
	a hard TeV $\gamma$-ray source and likely powers the latter as
	a TeV halo,
	we select 3 candidate TeV halos from the first Large High Altitude 
	Air Shower Observatory (LHAASO) catalog of \gr\ sources. 
	The corresponding pulsars, given by the positional coincidences and
	property similarities, are PSR J1958+2846, PSR J2028+3332, and
	PSR J1849$-$0001. We analyze the GeV $\gamma$-ray data 
	obtained with the Large Area Telescope (LAT) onboard 
	{\it the Fermi Gamma-ray Space Telescope} for the first two pulsars,
	as the last is \gr\ quiet. We remove the pulsed emissions of
	the pulsars from the source regions from timing analysis, 
	and determine that there are no residual GeV emissions in the regions 
	as any possible counterparts to the TeV sources. Considering the
	previous observational results for the source regions and
	comparing the two pulsars to Geminga (and Monogem), 
	the LHAASO-detected TeV sources are likely the pulsars' respective
	TeV halos. We find that the candidate and identified TeV halos, 
	including that of PSR~J1849$-$0001, have luminosites at 50\,TeV
	(estimated from the differential fluxes)
	approximately
	proportional to the spin-down energy $\dot{E}$ of the pulsars, and
	the ratios of the former to the latter are $\sim 6\times 10^{-4}$.
\end{abstract}


\section{Introduction}

Invoked by the detections of extended TeV emissions around
the Geminga and Monogem pulsars \citep{abe+17}, it has been suggested
that very-high-energy (VHE; $\geq 100$\,GeV) TeV halos could be a common 
phenomenon associated with
middle-age ($\sim$100\,kyr) pulsars \citep{lin+17}. How such TeV halos
are formed is under intense theoretical studies (e.g., 
\citealt{elm18, lg18, tp19, fby19, lyz19}, and also see \citealt{ml22}
and references therein); generally electrons/positrons emanated from a pulsar
have somehow slowly diffused, forming a region larger than a pulsar wind nebula (PWN;
see, e.g., \citealt{sud+19, gia+20})
and emitting the observed TeV photons. On the basis of current 
speculations,
$\sim$100 TeV halos may be detectable in our Galaxy 
(e.g., \citealt{lin+17, sud+19}), and studies of them
would help clarify the pulsars' contribution to cosmic electrons/positrons
(e.g., \citealt{mdd20}).

In our recent studies of the region of a middle-aged pulsar J0631$+$1036
\citep{zwx23},
which has a high positional coincidence with a TeV source likely detected 
with the High-Altitude Water Cherenkov (HAWC) Observatory 
as 3HWC~\hawc\ \citep{3hawc} and with the Large High Altitude Air Shower 
Observatory (LHAASO; \citealt{lhaaso19}) as 1LHAASO~\lasso\ \citep{cao+23}, 
no GeV \gr\ emission was found at the region in the data obtained with 
the Large Area 
Telescope (LAT) onboard {\it the Fermi Gamma-ray Space Telescope (Fermi)}.
The non-detection, which sets a constraint on the existence of a 
PWN associated with the pulsar, strongly suggests the TeV source as
a TeV halo powered by the pulsar, as PWNe detected at TeV energies
appear to have soft emissions \citep{hesspwn18} and most of them can be 
detectable at GeV energies with \fermi\ LAT (see, e.g., \citealt{3pc23};
Zheng et al., in preparation).
This possibility is supported by the
great similarities in the X-ray properties of the pulsar with Geminga and
the properties of the TeV source with the TeV halo of Geminga.
\begin{table*}
	\begin{center}
	\caption{Timing parameters derived for PSR J1958$+$2846 and PSR J2028$+$3332}
	\label{tab:ta}
	\begin{tabular}{lcccc}
        \hline
	\hline
    Source& Time range & $f$  & $f_1/10^{-12}$  & $f_2/10^{-22}$ \\
		& (MJD) & (Hz)       & (Hz s$^{-1}$)   & (Hz s$^{-2}$)   \\
	\hline
	J1958 & 54682--56540   & 3.44 & 3489744(9)     & $-$2.5133(1) \\
	        & 56540--56740 & 3.44382(7)  & $-$13(2)    &  2204(470)\\
	       & 56740--59540    & 3.44349004(5)    & $-$2.521(1) &  1.9(3) \\
       \hline
        J2028 & 54682--59991    & 5.65906764938(2)  & $-$0.1555548(9) & 0.000353(9)   \\ \hline
\end{tabular}
\begin{tablenotes}
	\footnotesize
	\item The frequency epoch is MJD 55555 for both pulsars.
\end{tablenotes}
\end{center}
\end{table*}

Following the studies, we have checked for potential TeV halos among
the sources in the current
VHE source catalogs, which include that reported by the High Energy 
Spectroscopy System (HESS) Galactic plane survey \citep{hess18}. 
As argued in \citet{zwx23}, sources were selected if they have
hard TeV emission (i.e., possibly having an energy spectrum peaking 
around $\sim$25\,TeV) and do not have obvious
supernova remnant (SNR) or PWN counterparts, but have a high positional 
coincidence with a pulsar. We found three such sources,
HESS J1849$-$000 (or 1LHAASO~J1848$-$0001u), 1LHAASO~J1959+2846u, and
1LHAASO~J2028+3352 that
are in possible association with PSR~J1849$-$0001, PSR~J1958+2846,
and PSR~J2028+3332 (hereafter J1849, J1958, and J2028, respectively).
However, the first one J1849 is an X-ray pulsar that is both
radio and \gr\ quiet \citep{got+11,bog+19}. We thus conducted analysis
of the \fermi\ LAT data similar to that in \citet{zwx23} for 
J1958 and J2028. These two pulsars are \gr\ bright, first discovered from
the \fermi\ LAT observations \citep{j19dis09,ple2012}, while J2028 is radio
quiet (\citealt{gri+21} and references therein).

In this paper, we report on the results from our analysis for the two pulsars.
The analysis and results are presented below in
Section~\ref{sec:obs}. Based on the results and properties of the TeV sources
(1LHAASO~J1959+2846u and 1LHAASO~J2028+3352) and pulsars, we suggest that 
the TeV sources are likely TeV halos of 
the pulsars. This discussion, including a summary for the possible properties
of the TeV halos, is presented in Section~\ref{sec:dis}.

\section{Data Analysis}
\label{sec:obs}

\subsection{LAT data and source model}
\label{sec:sm}

We used the $\gamma$-ray data collected with \fermi\ LAT \citep{atw+09}. 
The two regions of interest (RoIs) were set to have a size of 
$15^{\circ} \times 15^{\circ}$ with each centered at the positions 
of J1958 and J2028 respectively. The events in each RoI in the energy range 
of 0.1--500 GeV over the time period of from 2008 August 04 15:43:36 (UTC) 
to 2023 February 16 00:00:00 (UTC; approximately 14.5\,yr) were selected
from the latest \fermi\ Pass 8 database.
Following the recommendations of the LAT team\footnote{\footnotesize http://fermi.gsfc.nasa.gov/ssc/data/analysis/scitools/}, 
we excluded the events with quality flags of `bad' and zenith 
angles $\geq$ 90\arcdeg.

We used the \fermi\ LAT 12-year source catalog 
(4FGL-DR3; \citealt{4fgl-dr3}) to construct the source models.
The sources within 15\arcdeg\ of the pulsar targets in the catalog 
were included in the source models, and their spectral forms provided by
the catalog were used. 
[Since the \fermi\ LAT Fourth source catalog 
was just updated (4FGL-DR4; \citealt{bal+23}) 
while we were finishing this paper, we checked for possible differences
between the two releases. For the RoIs, there are no significant differences,
and for the regions of approximately 5\arcdeg\ around the pulsars, there
were no differences.]
The spectral model files gll\_iem\_v07.fits and 
iso\_P8R3\_SOURCE\_V3\_v1.txt were used as the background Galactic and 
extragalactic diffuse emissions respectively.  
\begin{figure*}
\centering
\includegraphics[width=0.47\textwidth]{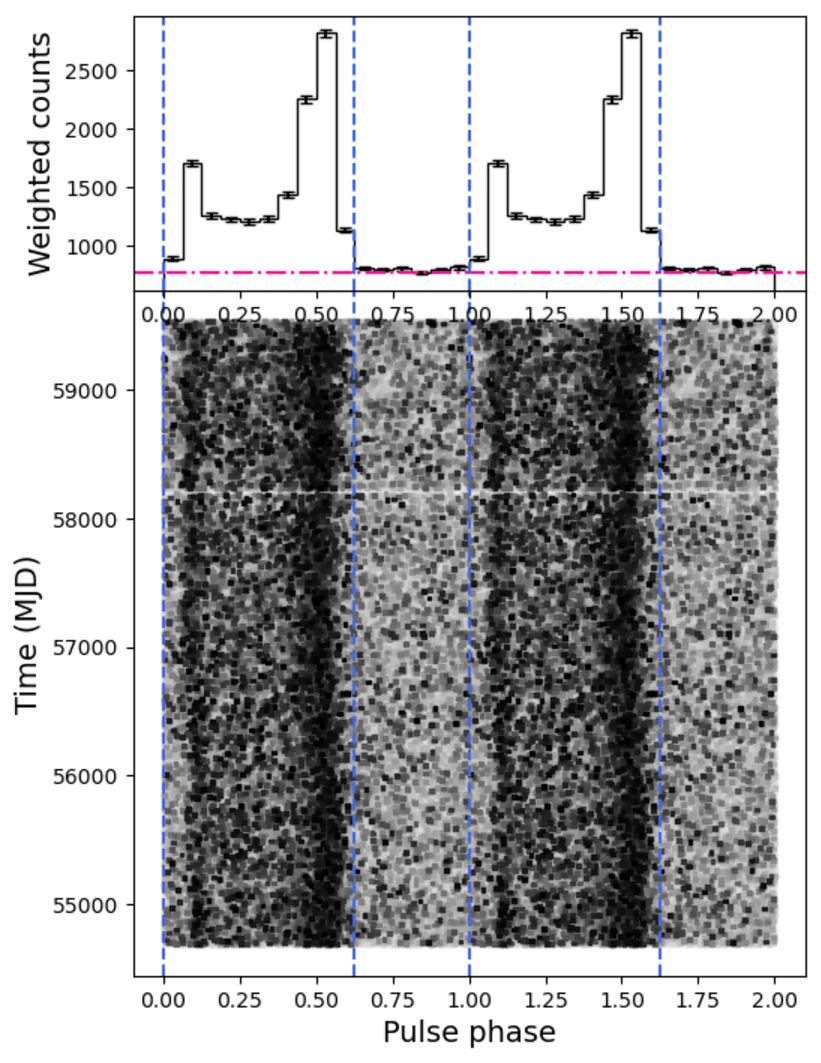}
\includegraphics[width=0.47\textwidth]{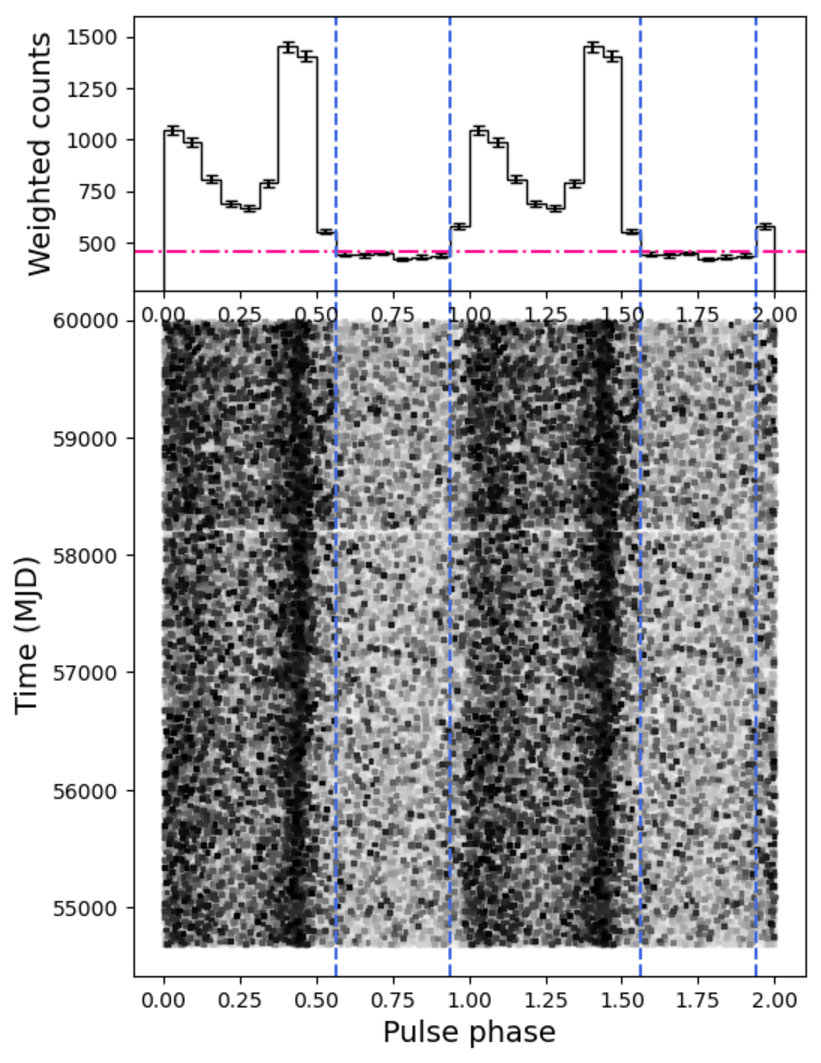}
	\caption{Phase-connected pulse profile ({\it top}) and two-dimensional
	phaseogram ({\it bottom}) constructed for J1958 ({\it left})
	and J2028 ({\it right}).
	Two spin cycles are shown for clarity. The onpulse and offpulse
	phase ranges are marked by dashed lines.}
\label{fig:pp}
\end{figure*}

\subsection{Timing analysis}
\label{sec:ta}

The two pulsars are bright in the LAT \gr\ band. We selected photons within an 
$1^{\circ}$ radius circular region centered at each of them
for the purpose of constructing their pulse profiles and thus determining the
offpulse phase ranges. 

\subsubsection{PSR J1958$+$2846}

We first tested to fold the photons with the ephemeris given in LAT 
Gamma-ray Pulsar Timing Models (GPTM) database\footnote{https://confluence.slac.stanford.edu/display/GLAMCOG/LAT+Gamma-ray+Pulsar+Timing+Models} \citep{ray+11}.
No clear pulse profile over the 14.5-yr time period could be obtained.

We then used the method described in \citet{xin+22}, in which
the photons during MJD 54682--56540 (the similar time period to that in LAT 
GPTM database) were folded according to the known ephemeris by using 
the \fermi\ TEMPO2 plug-in \citep{ehm06,hem06}.
An empirical template profile was constructed, based on which
the times of arrival (TOAs) for as many as possible sets 
of $\sim 200$ day LAT data were generated.
The template and TOAs were obtained using the maximum likelihood method 
described in \cite{ray+11}. We fitted the TOAs with TEMPO2 
by adding high-order frequency derivatives and updated the ephemeris.
However this ephemeris, whose main parameters $f$, $f_1$, and $f_2$
are given in Table~\ref{tab:ta}, could only cover the data before MJD 56540.

For the data after MJD 56540, we were able to find two template profiles 
during MJD 56540--56740 and MJD 56740--59540, while the same method as the
above was used. The updated ephemerides are given in Table~\ref{tab:ta}.

We folded the photons in the three time periods according to the ephemerides 
respectively to construct the pulse profiles. There were phase shifts 
between them, $\simeq 0.4375$ between the first and second and $\simeq 0.125$ 
between the first and the third.  
After making the corrections for the phase shifts, a pulse profile over 
nearly 13.3-yr was obtained (left panel of Figure~\ref{fig:pp}). 
Based on the profile, we defined phase 0.0--0.625 as the onpulse phase range 
and phase 0.625--1.0 as the offpulse phase range. 

\subsubsection{PSR J2028$+$3332}

For this pulsar, the selected photons 
could be easily folded according to the ephemeris given in LAT GPTM
database. A pulse profile over the whole LAT data time period was obtained,
while the ephemeris used is given in Table~\ref{tab:ta}.
Based on the pulse profile, which is shown in the right panel of
Figure~\ref{fig:pp}, we defined phase 0.0--0.5625 and 0.9375--1.0 
as the onpulse phase ranges and phase 0.5625--0.9375 as the offpulse phase 
range.
\begin{figure*}
	\centering
\includegraphics[width=0.47\textwidth]{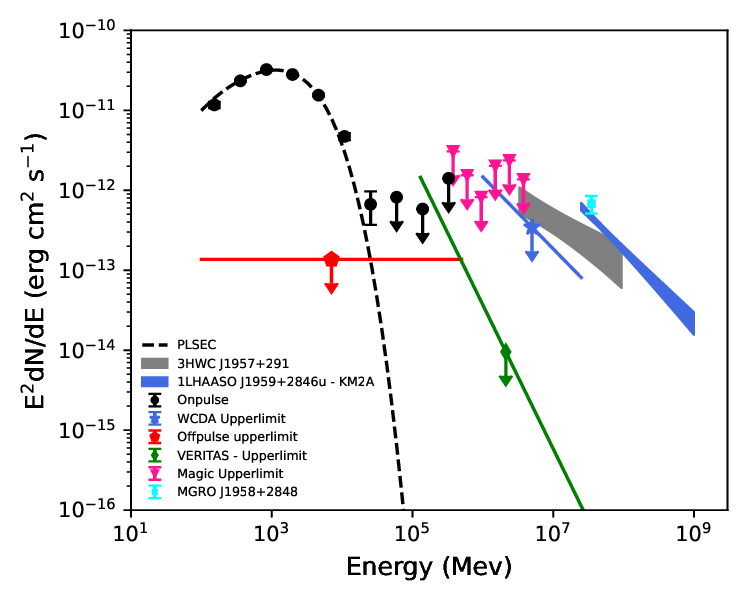}
\includegraphics[width=0.47\textwidth]{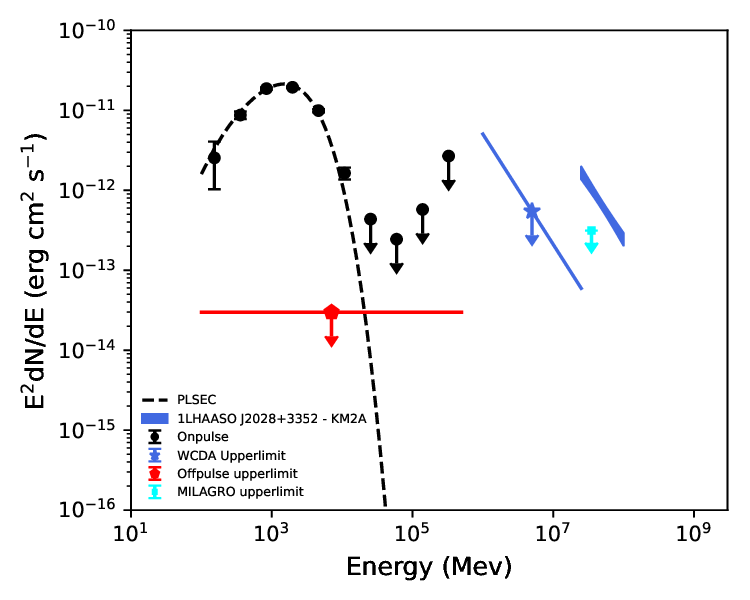}
	\caption{\gr\ spectra (black dots) of J1958 ({\it left}) and J2028 
	({\it right}) in 0.1--500\,GeV obtained from their onpulse data and the
	corresponding best-fit PLSEC models (dashed curves). From the offpulse
	data, the upper limits derived by assuming a power law with $\Gamma=2$ 
	are shown as the red lines.  The fluxes
	or flux upper limits from different VHE observations of
	the source regions are also shown, which include 
	in the {\it left} panel,
	1LHAASO J1959+2846u measured with the Kilometer Squared Array 
	(KM2A) and
	Water Cherenkov Detector Array (WCDA; \citealt{cao+23}), 
	MGRO J1958+2848 \citep{milagro09}, 
	VERITAS upper limits on J1958 \citep{arc+19}, and MAGIC upper limits
	on the PWN of J1958 \citep{fer+17}, and in the {\it right} panel 
	1LHAASO J2028+3352 and MGRO upper limit on J2028.
 \label{fig:spec}}
\end{figure*}

\subsection{Likelihood and spectral analysis}
\subsubsection{Onpulse data}
\label{sec:od}

We performed standard binned likelihood analysis to the 0.1--500\,GeV LAT data 
during the onpulse phase ranges of two pulsars determined above. The source 
models described in Section~\ref{sec:sm} were used.
The sources within $5^{\circ}$ from each of the pulsars were set to have 
free spectral parameters, while for the other sources in the source models, 
their spectral parameters were fixed at the values given in the catalog. 
The background normalizations were set as the free parameters.

For fitting the emissions of the pulsars,
we used a PLSuperExpCutoff4 (PLSEC) model \citep{4fgl-dr3}), $\frac{dN}{dE} = N_{0} (\frac{E}{E_{0}})^{-\Gamma - \frac{d}{2} \ln(\frac{E}{E_{0}}) - \frac{db}{6} \ln^{2}(\frac{E}{E_{0}}) - \frac{db^{2}}{24} \ln^{3}(\frac{E}{E_{0}})}$, 
where $\Gamma$ is the photon index, $d$ the local curvature 
at $E_{0}$, and $b$ a measure of the shape of the exponential 
cutoff and was fixed at $2/3$ (a characteristic value used for 
the $\gamma$-ray pulsars in 4FGL-DR3). From the likelihood analysis,
we obtained 
$\Gamma = 2.12 \pm 0.01$ and $d = 0.63 \pm 0.03$
($\Gamma = 2.140 \pm 0.004$ and $d = 1.109 \pm 0.007$) for J1958 (J2028). 
These values are close to those given in 4FGL-DR3. These results, as well 
as the corresponding TS values, are provided in Table \ref{tab:ts}.

We also obtained the flux measurements of the \gr\ emissions
of J1958 and J2028
from their onpulse data.  The number of the energy bins was set to be 10 
by evenly 
dividing the energy range of from 0.1 to 500\,GeV in logarithm.
Fluxes were obtained from the maximum likelihood analysis 
of the data in each energy bin.
In this analysis, the spectral normalizations 
of the sources within $5^{\circ}$ from each pulsar were set as free parameters,
and all other parameters of the sources in the source models were fixed 
at the values obtained from the above binned likelihood analysis.
When the TS value of a spectral data point was less than 4,
we used the 95\% flux upper limit instead.  The obtained spectra are shown 
in Figure~\ref{fig:spec}.

\subsubsection{Offpulse data}

We performed standard binned likelihood analysis to the 0.1--500 GeV LAT data 
during the offpulse phase ranges of the two pulsars. The parameter setups 
were the same as those in the analysis of the onpulse data 
(Section~\ref{sec:od}).  We assumed a power law for any emission at the position
of each pulsar, $dN/dE = N_{0} (E/1\ {\rm GeV})^{-\Gamma}$. No emissions were
detected. When we fixed 
$\Gamma=2$, the resulting TS values were $\sim$0 (Table~\ref{tab:ts}).

\begin{figure*}
	\centering
\includegraphics[width=0.47\textwidth]{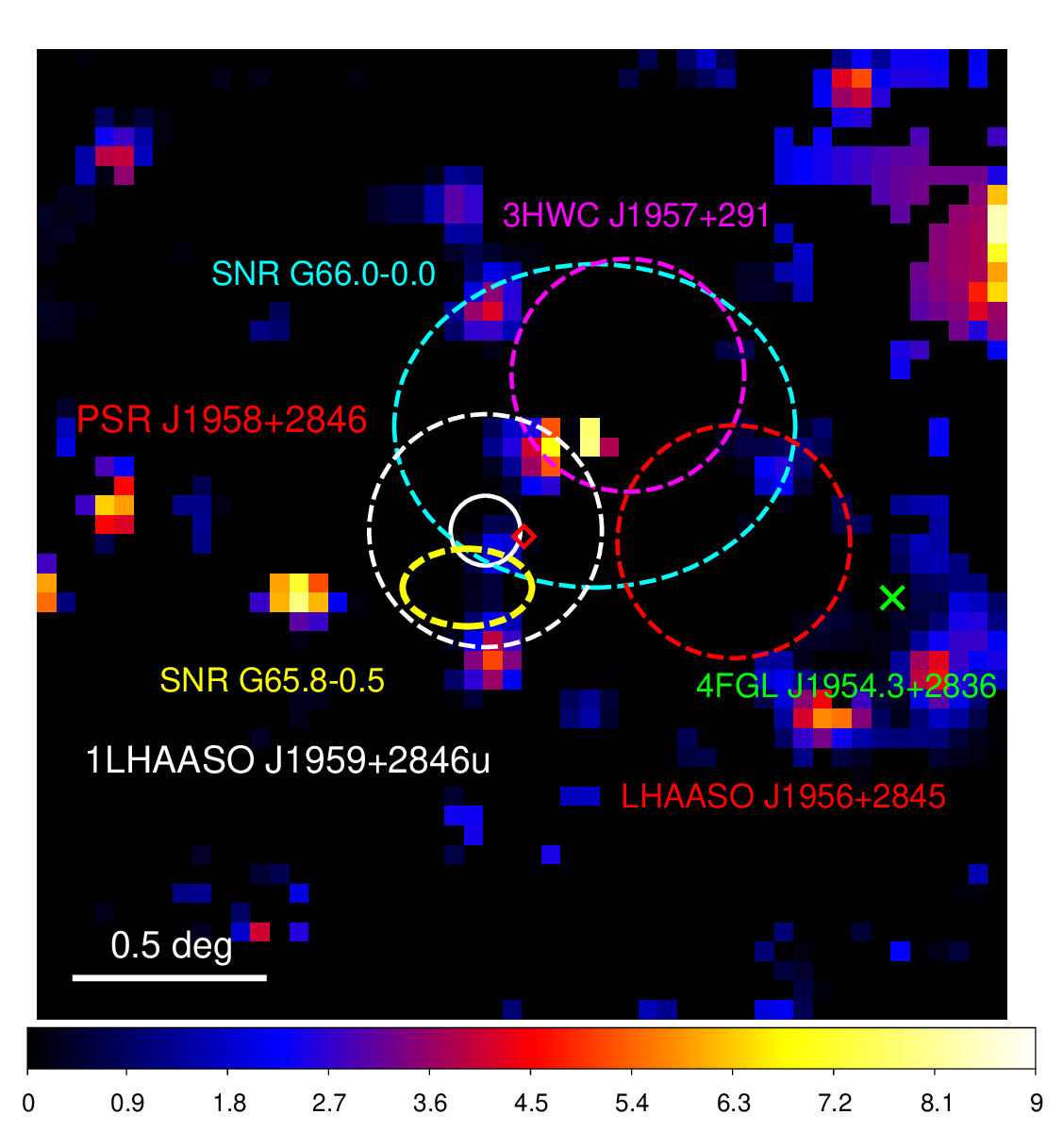}
\includegraphics[width=0.47\textwidth]{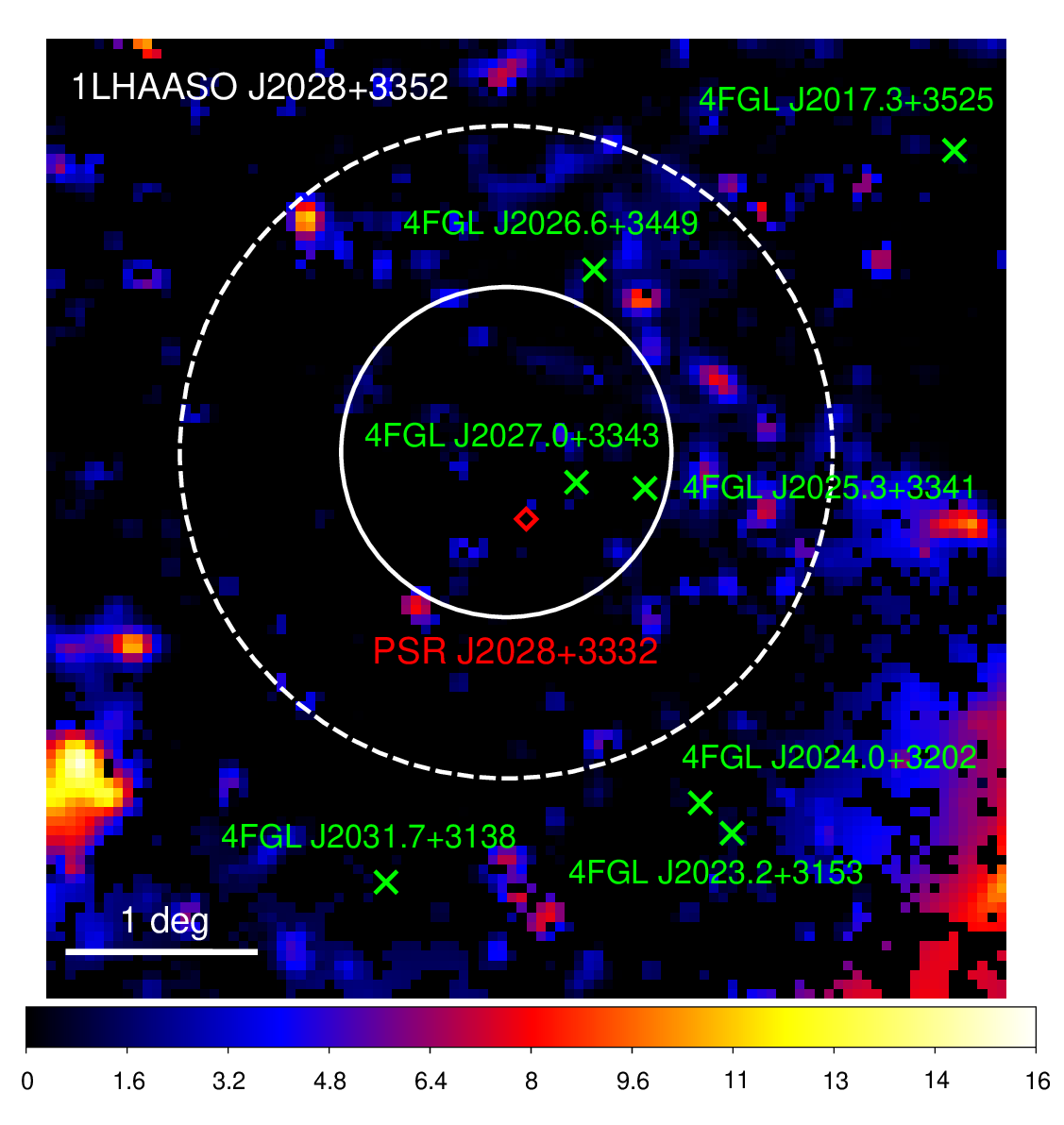}
	\caption{TS maps of the J1958 ({\it left}) and J2028 ({\it right})
	regions in 0.1--500\,GeV calculated for the offpulse phase ranges
	of the two pulsars. In the {\it left} panel, the positional error circle
	and extension of 1LHASSO J1959+2846u \citep{cao+23} are marked by white
	solid and dashed circles respectively, which are coincident with J1958's
	position (red diamond). Also shown are the positional error circles
	of 3HWC J1957+291 (pink dashed; \citealt{3hawc}) and LHAASO 
	J1956+2845 (red dashed; \citealt{cao+21}). In addition, two candidate
	SNRs, G65.8$-$0.5 and G66.0$-$0.0 (marked by yellow dashed 
	and cyan 
	dashed ellipses respectively), reported in \citet{sab+13} are 
	located in the region as well. In the {\it right} panel, the positional
	error circle and extension of 1HASSO J2028+3352 are marked as
	white solid and dashed circles respectively. The pulsar J2028, 
	along with
	two \fermi\ LAT sources, are located in the error circle.
 \label{fig:ts}}
\end{figure*}

To verify the non-detection results in the offpuse data, we calculated 
0.1--500\,GeV TS maps for the two pulsar regions. PWNe or TeV halos
may show extended weak emission at GeV energies (e.g., \citealt{dim+21}). 
However, as shown by the TS maps
(Figure~\ref{fig:ts}), no residual emissions are seen in each of the pulsar 
regions. We also tested to assume a uniform-disk model (with a radius
value varied upto 1\arcdeg) centered at each of the pulsars and perform
binned likelihood analysis. No extended emissions were detected, which confirms
the visual inspection results of the TS maps.

\section{Discussion}
\label{sec:dis}

By selecting TeV sources sharing the similarities with the TeV halo of Geminga,
i.e., having hard emission and being positionally coincident with only 
a middle-aged pulsar, we have found 1LHAASO~J1959+2846u and
1LHAASO~J2028+3352 as possible TeV halos. The corresponding pulsars J1958 and
J2028 both have bright \gr\ emissions in the \fermi\ LAT GeV band. We conducted
timing analysis of the LAT data to remove the pulsed emissions of the pulsars
from the source regions, and found that no GeV \gr\ emissions were detected
in each of the regions. The non-detections largely constrain the existence 
of PWNe 
associated with the pulsars, as we note that the spectral upper limits of
$\sim$10$^{-13}$\,erg\,s$^{-1}$\,cm$^{-2}$ (Figure~\ref{fig:spec}) 
set a luminosity limit on a PWN
at a distance of 2\,kpc down to $\sim$10$^{32}$\,erg\,s$^{-1}$, lower than
those of the known PWNe or candidate PWNe (e.g., \citealt{ack+11}). Combining
the non-detections of any GeV emissions with the hard emission property of 
the two TeV sources, it is likely that they are TeV halos of the two pulsars.

Below we discuss the different detection results for the source regions 
of the two 
pulsars in Section~\ref{sec:j19} \& \ref{sec:j20}, which help strengthen 
their likely associations
with the two TeV sources. In Section~\ref{sec:com}, we show the detailed
similarities of the two pulsars and their presumed TeV halos with the other
pulsar TeV halos, in which we include the pulsar J1849.

\subsection{PSR J1958+2846}
\label{sec:j19}

As shown in the left panel of Figure~\ref{fig:ts}, in addition to
1LHAASO~J1959+2846u that is detected with an extension size of 0\fdg29 in 
the energy range of $\geq$25\,TeV and has not been detected in 1--25\,TeV,
the Milagro gamma-ray observatory (MGRO) had a 4$\sigma$ detection at
the position of J1958 in 1--100\,TeV \citep{milagro09}, and HAWC and LHAASO
respectively reported two nearby sources 3HWC~J1957+291 \citep{3hawc}
and LHAASO~J1956+2845 \citep{cao+21}. The MGRO detection is consistent with
the LHAASO results (Figure~\ref{fig:spec}), but the latter two have
separation distances of 0\fdg54 and 0\fdg74 respectively. We suspect that
the latter two are individual sources, not likely in association, 
which can be verified by near-future updating observations.
In addition, the MAGIC telescopes searched for the PWN of J1958 \citep{fer+17}
and VERITAS telescopes searched for pulsed \gr\ emission of 
J1958 \citep{arc+19}, both at the energy of $\gtrsim$100\,GeV, but no emission
was detected (see Figure~\ref{fig:spec} for their upper limits).
Finally, we note that there are two candidate SNRs identified from an optical
H$\alpha$ survey \citep{sab+13}, which positionally overlap with the pulsar
or 1LHAASO~J1959+2846u. However very limited information is available for
these two candidate SNRs (e.g., \citealt{gre19}). We suggest they are not 
likely the counterpart
to the TeV source given the non-detection of any GeV \gr\ emissions in
the offpulse data.

\subsection{PSR J2028+3332}
\label{sec:j20}

The field of J2028 is rather clean, as 1LHAASO J2028+3352, having a positional
error circle of 0\fdg86 and an extension size of 1\fdg7, is the only known
VHE source reported (Figure~\ref{fig:ts}). The source has similar hard emission,
detected in 25--100\,TeV but not in 1--25\,TeV. There was a reported upper limit
on the pulsar from MGRO observations \citep{milagro09}, and the upper limit
is lower than that of the LHAASO measurements (Figure~\ref{fig:spec}). This
inconsistency is not understood, probably due to the simple assumption
of a point source with a fixed power-law spectral form in the MGRO data 
analysis.
We note that there are two
\fermi\ LAT sources also detected in the error circle, 4FGL~J2025.3+3341 and
4FGL~J2027.0+3343. Given in the LAT catalog, the first is identified as a 
blazar, thus not likely the counterpart to 1LHAASO J2028+3352, and the second 
is an unknown-type source with its emission described with a power law and
is faint (TS$\sim$40). The nature of this second GeV source remains to 
be investigated, helping clarify whether its presence is purely because 
of a coincidence due to the large error circle of the TeV source.

In addition, since J2028 is radio quiet, there is no distance estimation for 
the source.  We obtained its phase-averaged \gr\ flux, 
$\simeq$5.1$\times 10^{-11}$\,erg\,cm$^{-2}$\,s$^{-1}$ in 0.1--500\,GeV.
Given its spin-down energy $\dot{E}=3.5\times 10^{34}$\,erg\,s$^{-1}$, the
distance should be $\leq$2.4\,kpc. If we further consider an efficiency of
10\% for converting $\dot{E}$ to the \gr\ emission \citep{3pc23}, 
the distance would be 
$\simeq$0.76\,kpc. Therefore
the pulsar is likely close, which can explain the large extension size of 
1LHAASO J2028+3352 if we consider the latter as the putative TeV halo.

\begin{figure*}
	\centering
\includegraphics[width=0.47\textwidth]{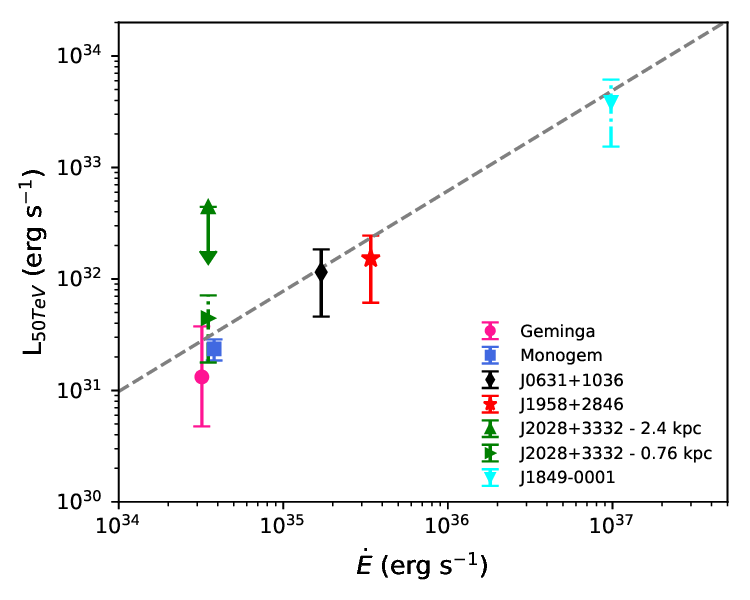}
\includegraphics[width=0.47\textwidth]{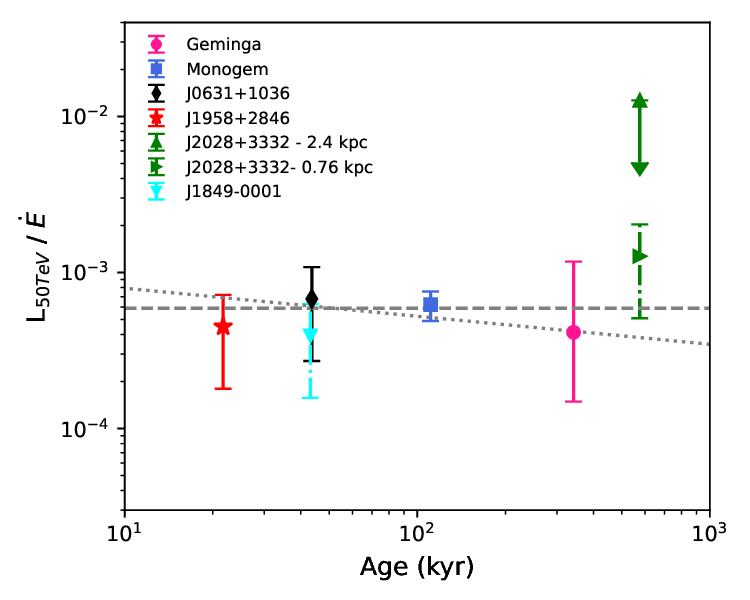}
	\caption{{\it Left} panel: Luminosity values at 50\,TeV from the LHAASO
	measurements of the (candidate) TeV halos versus $\dot{E}$ of 
	the pulsars. A relationship of $L_{\rm 50TeV} \simeq 2.5\dot{E}^{0.90}$
	(marked as the dashed line) can approximately fit the data points.
	{\it Right} panel: ratios of the luminosity values at 50\,TeV to 
	$\dot{E}$ of the pulsars, which shows that the ratios are around
	the value
	of $\simeq 5.9\times 10^{-4}$ (marked by the dashed line) despite the
	large age differences of the pulsars. In addition, an
	age-dependent relationship, $1.2\times 10^{-3}\tau_{\rm kyr}^{-0.18}$, 
	is also shown as the dotted line (see Section~\ref{sec:com}).}
	\label{fig:comp}
\end{figure*}

\subsection{Property comparison}
\label{sec:com}

We collected the properties of these two pulsars as well as Geminga, Monogem,
and PSR~J0631+1036, which are given in Table~\ref{tab:thalo}. In addition,
although J1849 is \gr\ quiet and we did not conduct analysis for
it, we also list it in the table because the corresponding TeV source
1LHAASO J1848--0001u shows similar hard emission. These pulsars have spin
periods of 0.2--0.4\,s except J1849 whose value is only $\simeq$0.04\,s
The characteristic ages spread from 20 to 600\,kyr. They are mostly X-ray
faint and have faint (or non-detectable) X-ray PWNe also 
except J1849 (considering its assumed distance of $\sim$7\,kpc;
\citealt{got+11}). Since they are all detected by LHAASO in 25--100\,TeV, 
we also list their fluxes at 50\,TeV \citep{cao+23} in Table~\ref{tab:thalo}.

As noted in \citet{zwx23}, the putative TeV halo of PSR~J0631+1036 has a flux
proportional to $\dot{E}$ of the pulsar when comparing to the corresponding
values of Geminga (where the HAWC measurements were used). We thus plot
the luminosity values at 50\,TeV of the (candidate) TeV halos given by
the LHAASO detections versus $\dot{E}$ of the pulsars in the left panel of
Figure~\ref{fig:comp}, where for the source distances (PSR J0631+1036 and J1958)
given by the radio timing we assign 30\% uncertainties, for J2028 the upper 
limit is 2.4\,kpc and we also assign a 30\% uncertainty on the possible 
distance of 0.76\,kpc, and for PSR J1849 its suggested distance of 7\,kpc
is assumed to have a 30\% uncertainty as well. As can be seen, there is
a possible correlation between the 50\,TeV luminosities ($L_{\rm 50TeV}$)
and $\dot{E}$, although
the distances suffer large uncertainties. We test to use the four pulsars
with relatively certain properties, Geminga, Monogem, PSR J0631+1036, and
J1958 and find a correlation of 
$L_{\rm 50TeV}\simeq (2.5\pm 1.7)\dot{E}^{0.90\pm0.01}$, where the Markov 
Chain Monte Carlo (MCMC) code {\tt emcee} \citep{fd+13} was used.
The index value 0.90 is close to 1, which does suggest the fluxes of
the TeV halos are proportional to $\dot{E}$. It is also interesting to note
that J1849 is along this correlation line.

Another way to show this possible correlation is to plot the ratios of
$L_{\rm 50TeV}$ to $\dot{E}$. In the right panel of Figure~\ref{fig:comp},
we plot the ratios versus ages ($\tau$) of the pulsars. Despite the
large ranges mainly caused by the distance uncertainties, we test to
fit the data points of the four pulsars (given above) with a constant $c$
and a function of $a\tau_{\rm kyr}^b$ (where $\tau_{\rm kyr}$ is $\tau$
in units of kyr), 
and obtain $c=(5.9\pm1.1)\times 10^{-4}$
and $a=(1.2^{+2.3}_{-0.8})\times 10^{-3}$, $b=-0.18\pm 0.26$, where for the 
latter the MCMC method was used.
As for the best-fits, the corresponding values of
$\chi^2$ over degree of freedom (DoF) are 0.4/3 and 1.5/2 respectively, 
the latter does not
provide a better fit and thus the brightnesses of
TeV halos are likely not in any close relation with ages of pulsars.
We conclude that the ratios are approximately in a range of 
$4.8$--$7.0\times 10^{-4}$. 

From the comparison analysis, we may summarize the general properties
of pulsar TeV halos and thus the method to identify them. The TeV sources
should have hard emission with spectrum peaks $\gtrsim$ 25\,TeV, which can
be used to be differentiated among PWNe \citep{zwx23}. The source regions 
are rather
clean, without any potential SNR or PWN counterparts known at energies of
X-rays or $\gamma$-rays, while middle-aged pulsars can be found positionally
coincident with the source regions. The pulsars are generally \gr\ bright and
X-ray faint, and may have faint X-ray PWNe. Then if considering the 
relationship built based on the limited cases (i.e., Figure~\ref{fig:comp}),
the TeV halos would have fluxes proportional to $\dot{E}$ of the pulsars,
and a typical value for the ratios of TeV luminosities (at a given energy)
over $\dot{E}$ would be probably in a range of 10$^{-4}$--10$^{-3}$.

\begin{table*}
	\begin{center}
	\caption{Binned Likelihood Analysis Results}
	\label{tab:ts}
	\begin{tabular}{lccccc}
        \hline
	\hline
		Source  & Phase & $F^{\ast}/10^{-8} $ & $\Gamma$  & $d$ & TS \\\hline
J1958 & Onpulse & $11.9\pm0.3$     & $2.12\pm0.01$  & $ 0.63\pm0.03$ & 17819 \\
     & Offpulse$^\dagger$ & $\leq$0.09  & 2 & ---             &  0.7 \\ \hline
J2028 & Onpulse &$4.65\pm0.04$   & $2.140\pm0.004$  & $ 1.109\pm0.007$ & 11464\\
   & Offpulse$^\dagger$ & $\leq$0.02    & 2 & ---   &  0.0 \\
\hline
\end{tabular}
\begin{tablenotes}
        \footnotesize
\item $^\ast$Fluxes in energy 0.1--500\,GeV are in units of photon\,s$^{-1}$\,cm$^{-2}$
\item $^\dagger$Fluxes are 95\% upper limits derived with $\Gamma=2$ assumed 
\end{tablenotes}
\end{center}
\end{table*}

\begin{table*}
        \begin{center}
		\caption{Information for pulsars and their (candidate) TeV halos}
        \label{tab:thalo}
        \begin{tabular}{lccccccccc}
        \hline
        \hline
		Pulsar &  $P$ & $\dot{P}$   & $\dot{E}/10^{35}$   & Distance      & Age    & $F^{PSR}_{X}/10^{-13}$  & $F^{PWN}_{X}/10^{-13}$  & $F_{50 {\rm TeV}}/10^{-13}$ & References \\
		   &  (s)            & $10^{-14}$  & (erg s$^{-1}$)         & (kpc)    & (kyr)  & (erg cm$^{-2}$ s$^{-1}$)    &  (erg cm$^{-2}$ s$^{-1}$)       &  (erg cm$^{-2}$ s$^{-1}$)  & \\
        \hline
		Geminga  & 0.24 & 1.10 & 0.32 & 0.25$^{+0.23}_{-0.08}$ & 342 & $7.51^{+0.07}_{-0.75}$ & 0.81 $\pm$ 0.03 &17.68 $\pm$ 0.60  & 1, 2 \\
		Monogem & 0.38 & 5.50 & 0.38 & 0.28$\pm$0.03 & 111 & 23.5 $\pm$ 0.4 &  $0.083^{+0.057}_{-0.044}$ &25.20 $\pm$ 0.92 & 1, 3 \\
        \hline
		J0631$+$1036 & 0.29 & 10.47 & 1.7 & 2.11 & 43.6 & $\leq0.09$ & ---  &2.16 $\pm$ 0.24  & 4, 5 \\
		J1958$+$2846 & 0.29 & 21.20 & 3.4 & 1.95 & 21.7 & 0.14 $\pm$ 0.04 & 0.04 $\pm$ 0.03  &3.36 $\pm$ 0.28 & 4, 6 \\
		J2028$+$3332 & 0.18 & 0.49 & 0.35 & $\leq$2.4 & 576 & 0.05 $\pm$ 0.03 & --- &6.44 $\pm$ 0.76  &  7, 8\\
		J1849$-$0001  & 0.04 & 1.42& 98 & 7 (?) & 43.1 & 38.0 $\pm$ 3.0 & 9.0 $\pm$ 2.0 &6.56 $\pm$ 0.40  & 9 \\ \hline
\end{tabular}
\begin{tablenotes}
        \footnotesize
\item References for distances and X-ray fluxes: 1) \citet{ver+12}; 2) \citet{pos+17}; 3) \citet{bpk16}; 4) \citet{man+05}; 5) \citet{ken+02}; 6) \citet{kar+12}; 7) this work; 8) \citet{mar+15}; 9) \citet{got+11}.
\end{tablenotes}
\end{center}
\end{table*}

\begin{acknowledgements}
This research is supported by 
the Basic Research Program of Yunnan Province
(No. 202201AS070005), the National Natural Science Foundation of
China (12273033), and the Original
Innovation Program of the Chinese Academy of Sciences (E085021002).
\end{acknowledgements}

\bibliographystyle{aasjournal}
\bibliography{thalo}

\end{document}